# Exploration of Optimizing FPGA-based Qubit Controller for Experiments on Superconducting Quantum Computing Hardware

Hans Johnson,[1,2,†] Silvia Zorzetti,[2] and Jafar Saniie [1]

[1] Embedded Computing and Signal Processing (ECASP) Research Laboratory (http://ecasp.ece.iit.edu)
Department of Electrical and Computer Engineering
Illinois Institute of Technology, Chicago IL, U.S.A.

[2] Superconducting Quantum Materials and Systems Center (SQMS),
Fermi National Accelerator Laboratory, Batavia, IL 60510, U.S.A.

[†] hjohnson1@hawk.iit.edu

*Abstract* — This work explores avenues and target areas for optimizing FPGA-based control hardware for experiments conducted on superconducting quantum computing systems and serves as an introduction to some of the current research at the intersection of classical and quantum computing hardware. With the promise of building larger-scale error-corrected quantum computers based on superconducting qubit architecture, innovations to room-temperature control electronics are needed to bring these quantum realizations to fruition. The QICK (Quantum Instrumentation Control Kit) is one leading experimental FPGA-based implementations. However, its integration into other experimental quantum computing architectures, especially those using superconducting radiofrequency (SRF) cavities, is largely unexplored. We identify some key target areas for optimizing control electronics for superconducting qubit architectures and provide some preliminary results to the resolution of a control pulse waveform. With optimizations targeted at 3D superconducting qubit setups, we hope to bring to light some of the requirements in classical computational methodologies to bring out the full potential of this quantum computing architecture, and to convey the excitement of progress in this research.

## I. Introduction

Quantum computing is a groundbreaking approach to computing that will shift how we solve some of the world's most technologically complex problems. Quantum computing relies on exploiting principles of quantum mechanics to easily overcome significant computational challenges currently presented to the most powerful computing systems in the world. Unlike classical computers, which use bits to store and manipulate information, quantum computers utilize quantum bits – also known as qubits – which can exist in multiple states at the same time. This is a physical property known as superposition, and it allows quantum computers to perform calculations much faster and more efficiently than classical computers.

The development of quantum systems and algorithms has the potential to solve complex problems that are currently intractable with classical computing methods. These applications include (but are not limited to) optimization problems, drug discovery utilizing quantum chemistry, finance, material science, simulations, dark matter research, cryptography, artificial intelligence, optimizing power grid networks and the integration of smart-grid technologies, and even designing more efficient computer chips [1–3]. However, building a practical quantum computer is a significant scientific and engineering undertaking. Quantum instruments are highly sensitive to their environment and can quickly lose their quantum properties; a process known as decoherence. Overcoming this challenge has made up a large portion of quantum system research in the past decade, and researchers are exploring a variety of approaches that include error-correcting codes, materials innovations, and improving control systems over quantum technology.

One of the more significant innovations in quantum systems has been the use of superconducting qubits and SRF (Superconducting Radio Frequency) cavities, an architecture known as 3D quantum processing units (QPUs). This is the primary design being explored at the Superconducting Quantum Materials and Systems Center (SQMS) at Fermi National Accelerator Laboratory (Fermilab). This technology has demonstrated supreme coherence times versus other quantum computing architectures [5] such as solid-state, ion trap, or photonic qubits. In 3D QPUs, superconducting qubits are the fundamental element of these circuits, while SRF cavities are used to control and manipulate the qubits. SRF cavities provide a highly stable and low-noise environment for qubits to operate in, which is crucial for maintaining coherence and minimizing errors in quantum experiments. They are also capable of generating extremely precise and tunable microwave signals which are used to control qubits.

The states of 3D QPU systems are typically manipulated by feeding precise radio frequency (RF) pulses through a control line. The control pulses interact with the quantum elements of the system, which in turn manipulate the pulses with new information about the system upon readout. State-of-the-art room temperature control electronics are necessary to orchestrate the control and readout process [4]. The frequencies of superconducting quantum systems typically lie within the 1 – 10 GHz bandwidth [5], which means any resonance frequencies must lie within the same band.

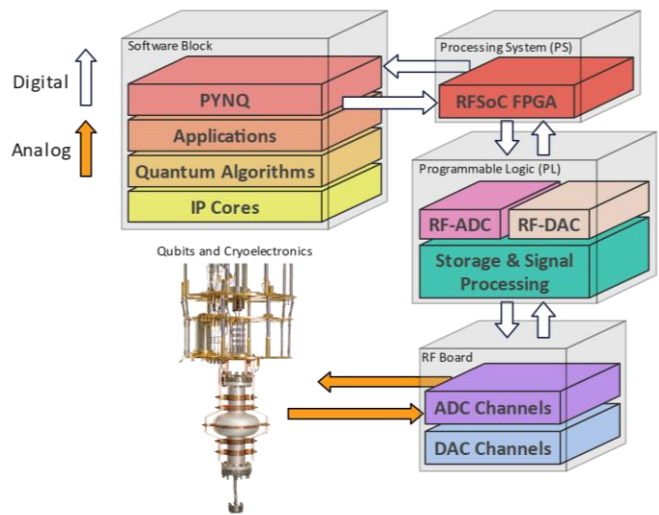

Fig. 1. 3D QPU Block Diagram Integrated with the QICK System

This has fueled the exploration of FPGA (Field Programmable Gate Array) technology as a controls hardware framework for Fermilab's quantum systems. Recently, Fermilab has developed its own qubit control module known as the QICK (Quantum Instrumentation Control Kit), which is a publicly available open-source project [6]. A general system architecture block diagram with the integrated QICK for a quantum system with superconducting qubits and a single-mode SRF cavity can be seen in Fig. 1. It is a hardware-agnostic implementation that can be used by a wide range of quantum architectures with unparalleled scalability. It has provided researchers with a powerful tool for controlling and manipulating superconducting qubits and has been adopted by several institutions and labs specializing in qubit research. Though state-of-the-art Arbitrary Waveform Generators (AWGs) could achieve waveforms that fit the requirements, the ever-changing character of quantum information science and research at SQMS requires adaptable and reconfigurable hardware as new experimental challenges are presented. Thus, similarly malleable electronic hardware is preferred to static hardware architectures.

Though the QICK was developed by the Scientific Computing Division group at Fermilab, the integration into SQMS's hardware stack is not one hundred percent optimized for their specific quantum systems. There are several areas available for optimization in this area, one of which is the improvement of control pulse fidelity for SQMS's 3D SRF cavities. This work addresses one small area applicable to SQMS's quantum systems with the QICK and introduces other areas for current and future research.

Section II describes the system design in terms of the experimental quantum computing architectures specifically relevant to this work as well as the FPGA-based controls hardware, firmware, and software. It also describes some vital background knowledge for understanding the context of the system architecture. Some preliminary results for one avenue of current research in control hardware optimization are shown in section III. Inferences and current research are discussed in section IV, and concluding remarks are put in section V followed by acknowledgments in section VI.

## II. SYSTEM DESIGN

There are a few mainstream quantum computing architectures in use today, each of which has its own approach to the creation and manipulation of qubits. The four most widely used are listed below in the order of their popularity:

1. Superconducting qubits
2. Ion Trap
3. Photonic qubits
4. Topological qubits

However, due to the experimental nature of quantum computing, it is impractical at this point to try and prove the supremacy of one architecture over another. Still, it appears superconducting qubits boast the longest coherence times for most experiments [7, 8]. Besides the long coherence times, superconducting qubits are preferred to many other architectures because they can be fabricated using existing semiconductor technology.

Control hardware deals with the intersection of information in the classical and quantum worlds. Classical control data must be extremely high quality and precise to enable both large coherent quantum states and to deal with the manipulation and readout of these quantum states [6]. As of now, state-of-the-art RFSoC FPGAs meet all the essential requirements for quantum control hardware and will continue to be an integral part of research in the field.

### A. Quantum Computing Architectures

There exist multiple experimental setups in quantum computing that have proven capable and promising results. For some of the larger-scale quantum computers available, such as those from IBM or Google [9], superconducting qubits are their main explored architecture. Describing all the different types of qubits in quantum computing systems escapes the main purpose of this work, but there are many outstanding papers on the subject such as [10].

#### 1) 3D QPUs

Describing 3D QPU architecture made from superconducting qubits coupled with SRF cavities is also largely beyond the scope of this paper, though the necessary information for understanding how the control hardware interacts with the quantum computing system is summarized in this section. Details of the 3D QPU used in research can be found in [4].

SRF cavities originally are from research in the field of particle accelerators as microwave cavities to propel particle beams. Due to their high-quality factor ($Q_0$), a measurement of how well the resonator stores energy, this results in long photon decay times which equates to longer coherence times of quantum bits of several orders of magnitude over current state-of-the-art architectures [4, 5]. In 3D QPU architecture, a transmon qubit is placed inside the cavity. These cavities are typically made of niobium (Nb), and a picture of them can be found in Fig. 2.

When coupled with superconducting qubits, the relaxation time of an SRF cavity refers to the time it takes for the qubit to return to its ground state after it has been excited by a microwave pulse. The relaxation time can be split into two different phenomena: T1 and T2 relaxation times.

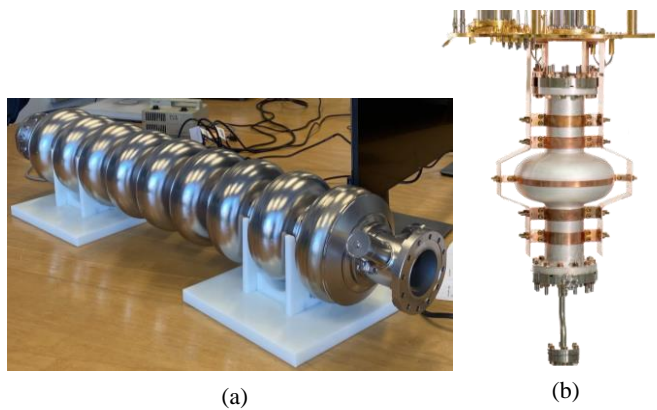

Fig. 2. SRF cavities: a) Multimode SRF Cavity, b) Single-mode SRF Cavity attached to the bottom of a Superconducting Quantum Computer

T1 relaxation time, also called longitudinal relaxation time, refers to the time it takes for the qubit to return to the ground state after initially being excited. It is especially important for determining the coherence of a qubit. T2 relaxation time, also called transverse relaxation time, refers to the time it takes for the qubit to lose coherence after it has been put into a superposition state by a microwave pulse. T2 relaxation time is an especially important parameter for determining the fidelity of quantum gates and implementing error correction schemas. In general, longer relaxation times lead to longer coherence times and higher fidelity quantum gates, which are essential to building practical quantum computers.

The T1 and T2 parameters are important considerations when designing control hardware for 3D QPUs. Longer relaxation times can lead to longer coherence times, but longer relaxation times can also lead to longer measurement times, longer gate times, and potentially more errors due to environmental noise with those longer timescales. One way to minimize these effects is to optimize the control hardware, such as using pulse shaping techniques to help reduce the time required for performing quantum gates and measurements. This would allow for the system to operate within the coherence times provided by the SRF cavity. Another approach is to use techniques like dynamic decoupling or continuous error correction with control hardware. By continuously monitoring and correcting for errors in real time, the system can maintain high-fidelity quantum operations even in the presence of longer relaxation times.

*2) Coherence Time*

Common to every quantum computer architecture is a phenomenon known as coherence. Coherence is an intrinsic property of qubits and represents how long a qubit can maintain its quantum state before being affected by noise or what's known as decoherence. The quantum property known as superposition allows a qubit to exist in a combination of both 0 and 1 simultaneously. However, when decoherence occurs, the system's quantum properties degrade due to its interactions with the environment. Decoherence is – if not the most – significant challenge in quantum computing, as it places limits on the time and conditions under which quantum computers can perform calculations.

Coherence times for superconducting qubits typically range from microseconds to hundreds of microseconds depending on the qubit design and quality of the fabrication process. This timing is a vital consideration for control hardware design and implementation because the coherence time represents the window of time one can perform quantum calculations and expect valid results. Control signals typically come in the form of pulses, and these pulses are associated with gate operations. The duration of these control pulses affects the fidelity of the operations, and for superconducting qubits typical control pulses are on the order of nanoseconds to tens of nanoseconds depending on the gate operation and qubit architecture. This means that a user will want to fit as many of these nanosecond control pulses in the microsecond coherence window before the quantum properties degrade.

*B. Controls Hardware and Firmware*

Across all quantum computing architectures, the ability to synthesize a large number of control signals with extreme accuracy and precision along with measuring the state of qubits and performing real-time feedback is essential. However, the equipment required for controlling tens of qubits and beyond is typically not affordable to academic labs and small startups. The RFSoC (Radio Frequency System-on-Chip) FPGA-based control system developed in the QICK supports experimental quantum computing architectures and is much more affordable to academic labs and small startups [6].

The QICK platform was implemented first on the Xilinx Zynq UltraScale+ RFSoC ZCU111 FPGA board with a custom RF board from Fermilab. However, since the release of the Gen 3 RFSoC FPGAs from Xilinx, most developments and implementations of the QICK are now done on the Zynq UltraScale+ RFSoC ZCU216 board along with Xilinx's XM655 breakout card included in the kit. A picture of the ZCU216 coupled with the XM655 breakout card can be seen in Fig. 3, and the specifications of the board and breakout card are highlighted in the block diagram in Fig. 4.

At the top level, the control pulses are calibrated using a software user interface known as PYNQ (Python for Zynq)

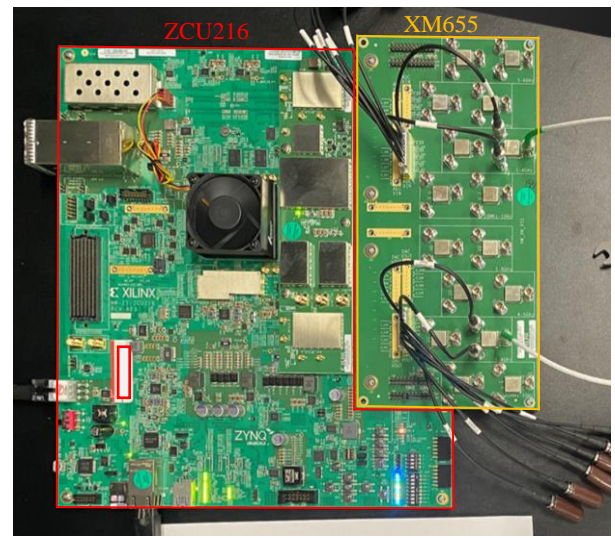

Fig. 3. ZCU216 UltraScale+ RFSoC with XM655 Breakout Card

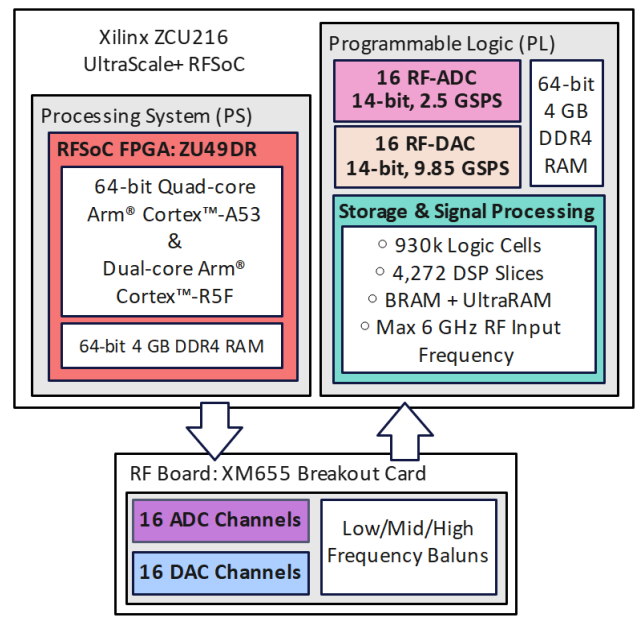

Fig. 4. FPGA-based Controls Electronics Block Diagram

into the QICK. As highlighted in Fig. 1, this data in software gets translated to FPGA-level instructions in the processing system and are then passed along to the programmable logic. From here, the system will then interact with the qubits in the fridges with a precise control signal via high-quality SMA cables, and the data is sent back through the stack.

It is important to note that all of the control electronics are managed at room temperature, and the signals directly to and from the quantum hardware are exclusively analog while all other computations and signal manipulations in the stack happen digitally.

*C. Software*

PYNQ is a framework designed to enable the use of Python for interacting with both the Processing System (PS) and the Programmable Logic (PL) of a Zynq SoC or FPGA with an ARM-based processing system. It provides an interface that allows users to develop and run Python code on the ARM processor in the PS, while also being able to control and communicate with the PL. PYNQ allows for the use of Python libraries and functions to interact with hardware and allows for the implantation of custom IP cores, accelerators, and/or peripherals. This is extremely useful for developers and programmers who need to access hardware/firmware but are unfamiliar with the electronics background to do so on traditional FPGAs.

The current use of the QICK system has been developed with PYNQ capabilities. Most of the open-source project is focused on optimizing use through PYNQ instead of actual firmware development. One reason for this is there is a significant need for the optimization of control pulses and quantum algorithms that can be solely developed and tested using PYNQ with the QICK. An example of what a basic control pulse for simulation may look like is found in Fig. 5. It exhibits a sinusoid contained in a gaussian envelope characterized by a flat top.

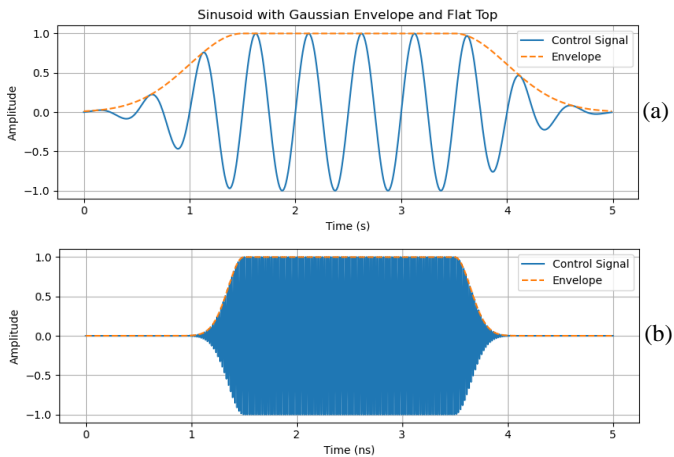

Fig. 5. Simplified Control Pulse: a) 2Hz Pulse, b) 5GHz Pulse used to simulate control pulse

*D. Control Pulse Fidelity and Implementation*

Control pulse fidelity is a metric of how accurately a control pulse can be generated and used is crucial for superconducting qubits and 3D QPUs because it affects the overall performance and coherence of the quantum system. High-fidelity control pulses enable accurate manipulation of qubit states and help in maintaining quantum states during quantum operations, which is essential for error reduction and achieving higher-quality quantum computations. Pulse shaping refers to the process of modifying the shape of a signal pulse to achieve specific desired characteristics or optimize performance in a control system. Precise pulse shaping helps minimize the leakage of quantum states into unwanted energy levels and reduces crosstalk between neighboring qubits, which can result in decoherence and other errors.

Though the QICK already employs brilliant pulse-shaping techniques that account for the limitations of the hardware, there are some areas that allow for different approaches and implementations that fit within the hardware constraints. The QICK system employs an FIR (Finite Impulse Response) filter for upsampling and downsampling in its control plane, which helps to generate high-fidelity control pulses for superconducting qubits. Pulse shaping techniques are used to reduce the impact of distortions introduced by the control system, such as frequency response and nonlinearities. It is easily reconfigurable and implemented in the digital domain to send signals to the DAC and the RF front-end to generate the analog control signals that manipulate the qubits [6].

Versus an FIR filter, a CIC (Cascaded Integrator-Comb) filter implementation can further mitigate errors introduced by the DAC and ADC in the control generation and acquisition process at the cost of little to no memory resources. This is because a CIC filter can efficiently interpolate a control pulse without requiring multipliers, as seen in the abstraction in Fig 6 below. This is especially important in a quantum system where multiple qubits need to be controlled simultaneously at the nanosecond scale.

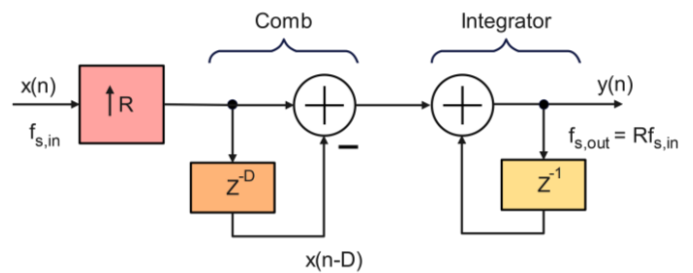

Fig. 6. Single Stage CIC Filter Structure Used in Interpolation

The biggest takeaway from Fig. 6 is that there are no multipliers required for the implementation of a CIC filter. This makes them an attractive choice for FPGAs (especially RFSoC) where resource utilization is a concern. They also exhibit low power consumption as well as high interpolation factors. It is also common to use a CIC filter in conjunction with an FIR filter to avoid deficits like a non-flat passband and limited stopband attenuation. This is because the passband in a CIC filter is not perfectly flat and the gain varies across the passband, causing distortion. Additionally, CIC filters have limited stopband attenuation relative to FIR filters which may result in unwanted frequency components in the filtered signal. In the case of using both a CIC and FIR filter, the CIC filter handles the initial interpolation, and the FIR filter refines the signal quality by compensating for the aforementioned deficits.

This is especially important with SRF cavities because the resonant frequencies of the SRF cavities in 3D QPU are sensitive to the control pulse shape, frequency, and phase. The CIC filter helps ensure that the desired frequency components are preserved while suppressing unwanted frequencies, which in turn allows for better control over qubit operations.

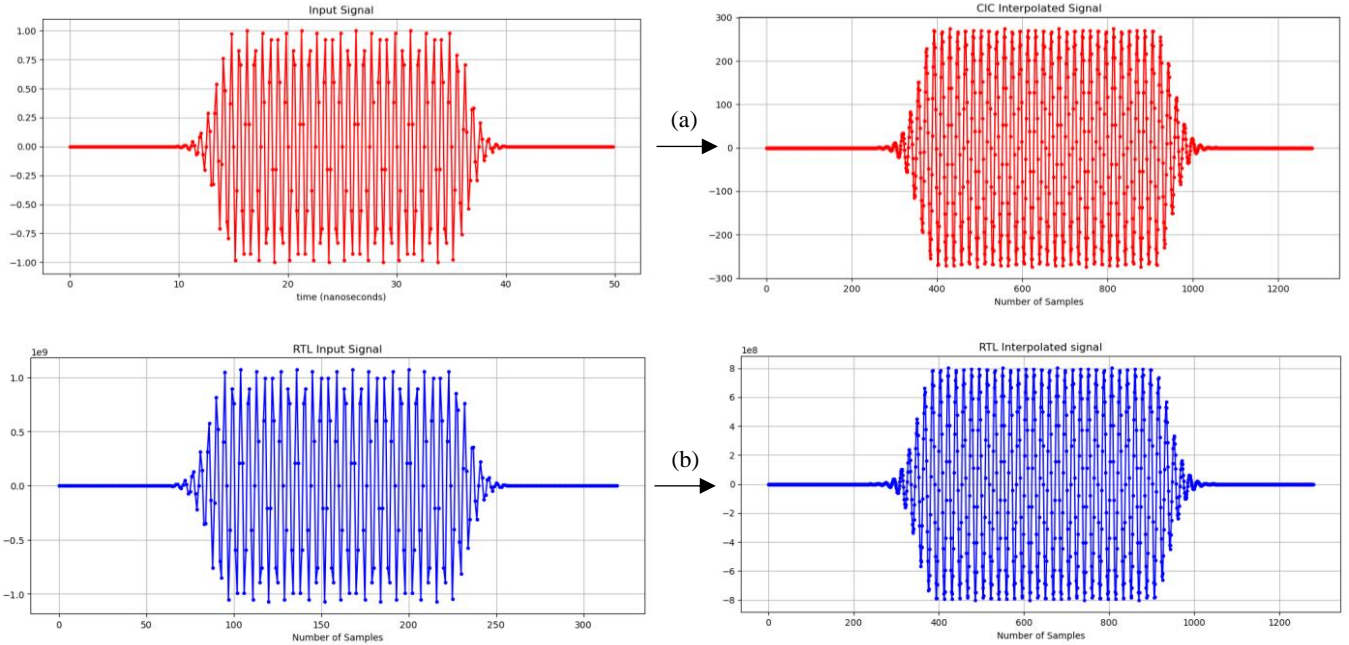

Fig. 7. Control Signals: a) Input Signal to x4 CIC Interpolated Signal using Python, b) RTL Simulation of 32-bit Fixed-Point representation of Input Signal to x4 CIC Interpolation using Vivado CIC Compiler IP (Verilog)

## III. RESULTS

Using Python, a simulation pulse was created for testing the functionality of the CIC filter designed to fit the specifications of the control hardware. This means that a pulse similar to Fig. 5 (b) was designed with a simple control pulse's parameters in mind in terms of frequency, amplitude, timing, and fit to the ZCU216's DAC sampling rate of 6.4 GSa/s. The control pulses need to be in the 4 to 6 GHz range which means the Nyquist rate is in the 8 to 12 GSa/s range, exceeding the requirements of the DAC. However, this is not a problem in actual implementations with the QICK due to the construction of the signal generator block and use of table memory and highly parallelized DDS [6]. Nonetheless, the CIC filter implementation can be used to interpolate control signals even after current pulse shaping processes while utilizing minimum resources to achieve a higher overall control pulse fidelity. A simulated pulse at 5 GHz and 6.4 GSa/s was used because the effects of undersampling and the benefits of interpolation can be observed easily.

Fig. 7 displays two of the simulation profiles used for the results, where Fig. 7 (a) uses data generated in Python as well as a custom CIC filter for x4 interpolation, and Fig. 7 (b) uses the specifications of the CIC filter built in Python to customize Xilinx's CIC Compiler IP block in Vivado to perform x4 interpolation over the data. The data was converted from floating-point to 32-bit fixed-point due to the specifications of the CIC Compiler IP block. This data conversion did result in a loss of some precision due to the limited resolution of the fixed-point format, but the loss was negligible in this test as observed in the near-identical interpolation graphs in Fig. 7. The compiler block was also instructed to use the DSP48E Dedicated Signal Processing (DSP) block available to Xilinx FPGAs. These DSP blocks are designed to efficiently perform complex arithmetic at super-fast speeds.

The RTL simulation was a success to prove the functionality of the actual build of the CIC block. However, implementing this block with the actual QICK firmware stack is a necessary step to prove the CIC filter's utility. Once that is verified, we can then test the implementation with actual qubit experiments on the quantum systems at SQMS.

## IV. DISCUSSION AND CURRENT RESEARCH

The interpolation of the signals was a success at x4 interpolation, and further work is required to observe higher interpolation rates with the same stopband attenuation. The filter was also adjusted to work optimally in the 4 to 6 GHz range. This is a very useful implementation because it does not require an overhaul of the current firmware, just the insertion of an already existing Vivado IP block.

This is one of many directions that optimizing the control electronics can go in for superconducting qubits, especially with implementations using SRF cavities. From a pure hardware standpoint, the ability to control many qubits with a multi-board FPGA stack is an interesting problem that will likely be relevant in the near future, as well as multiplexing qubits through the control board. Further alignment of coherence times in SRF cavities with optimized firmware will require other applications of pulse shaping techniques and fidelity/resolution innovations as well. Due to the experimental nature of SRF cavities, we must also take into consideration what we can do with optimizing gate errors, minimizing circuit depth for room-temp electronics, low-noise and high-bandwidth adjustments, integration and control of multiple qubits, reducing gate times through optimization and time required to perform quantum gates, and even new SoC FPGA board designs that better fit the requirements for 3D QPU.

## V. CONCLUSION

In this paper, one application was explored for optimizing the control hardware stack for superconducting qubit quantum systems. This implementation is just scratching the surface of what needs to be explored in the near future, but serves as a good introduction into some of the complexities of the control hardware in quantum computing. This multi-disciplinary research is in the beginning stages, but the promise for the use of SRF cavities coupled with superconducting qubits shows great promise in quantum computing technology. In conclusion, this work hopes to pioneer the foundation for a transformative leap in qubit control, and help in the realization of quantum computing technology.

## VI. ACKNOWLEDGMENTS

This material is based upon work supported by the U.S. Department of Energy, Office of Science, Office of Workforce Development for Teachers and Scientists, and Office of Science Graduate Student Research (SCGSR) program. The SCGSR program is administered by the Oak Ridge Institute for Science and Education for the DOE under contract number DE-SC0014664. This material is also based on work supported by the U.S. Department of Energy, Office of Science, National Quantum Information Science Research Centers, Fermilab's Superconducting Quantum Materials and Systems Center (SQMS), and Fermilab's Scientific Computing Division under contract No. DE-AC02-07CH11359. The authors thank the team and collaborators who developed the QICK board at Fermilab's Scientific Computing Division, notably Leandro Stefanazzi and Gustavo Cancelo for helping to understand the hardware and firmware design of the QICK system.